\documentstyle[graphicx,preprint,aps,epsfig]{revtex}

\begin{document}
\preprint{ SPIN-1999/25, UG-1999/42 }
\title{Confinement and the AdS/CFT Correspondence}
\author{David S. Berman$^1$\footnote{d.berman@phys.rug.nl} and
Maulik K. Parikh$^2$\footnote{m.parikh@phys.uu.nl} }
\address{${}^1$Institute for Theoretical Physics, University of Groningen, \\
Nijenborgh 4, 9747 AG Groningen, The Netherlands \\
${}^2$Spinoza Institute, University of Utrecht,\\
P. O. Box 80 195, 3508 TD Utrecht, The Netherlands}

\maketitle

\begin{abstract}
We study the thermodynamics of the confined and unconfined phases of
${\cal N} = 4$ Yang-Mills in finite volume and at large N 
using the AdS/CFT correspondence. 
We discuss the necessary conditions for a smooth phase crossover
and obtain an N-dependent curve for the phase boundary.

\end{abstract}

\def	\beq	{\begin{equation}} 
\def	\eeq	{\end{equation}}
\def	\lf	{\left (} 
\def	\rt	{\right )}
\def	\a	{\alpha} 
\def	\comma	{\; , \; \;} 

\section{Introduction}

It was realized many years ago \cite{planar} that in the large N limit of
Yang-Mills theory a remarkable simplification takes place: the physics is
dominated by ``planar'' graphs, Feynman diagrams with no line-crossings.
In this limit, the gauge theory ought to be described by a ``QCD string,''
and it was a hope that such a simplification might shed light on some of 
the mysteries of nonabelian gauge theories, notably the
puzzle of confinement.

The AdS/CFT correspondence \cite{adscft} 
makes explicit this relation between gauge fields and strings.
Specifically, the correspondence says that IIB string theory in a
background of five-dimensional anti-de Sitter space times a
five-sphere is dual to the large N limit of ${\cal N} = 4$
supersymmetric Yang-Mills theory in four dimensions. This is a conformal
theory with no confinement; however, the thermal theory in finite volume
could still have a confined phase.

An important ingredient in the correspondence is the principle of holography
\cite{hologerard,hololenny}, the notion that the physics of a gravitational
theory is dual to a different theory in one lower dimension.
Conversely, given the dual theory on a boundary, we must consider all
the possible bulk manifolds whose boundaries have the same intrinsic
geometry as the
background of the dual theory \cite{holographyads,thermalphase}.
For thermal super Yang-Mills on $S^3$, there 
are at least two distinct Einstein manifolds with the requisite boundary 
geometry: thermal anti-de Sitter space 
and a Schwarzschild-like black hole in AdS
\cite{hawkingpage}. These classical
bulk solutions are a sort of master field of the gauge theory. The distinct
geometries are interpreted in the gauge theory as 
different phases in the strong 't Hooft coupling limit;
the thermodynamics of the black hole
corresponds to the thermodynamics of strong-coupling SYM in the unconfined
phase while thermal AdS is seen as dual to 
the confined phase of the gauge theory.

In this paper, we investigate the thermodynamics
of the different phases of super Yang-Mills at finite volume on $S^3$ using the
AdS/CFT correspondence. In particular, we examine the conditions for
phase change in a microcanonical framework. Formally, a phase transition
cannot occur in finite volume at finite N. However, a crossover between
these qualitatively different phases can still occur when their weights
in the partition sum are the same. In the dual picture,
the black hole dominates the path integral when the
horizon is large compared to the inverse AdS curvature, and the thermal 
AdS geometry dominates for sufficiently low temperatures. The 
crossover between these two geometries is known as the Hawking-Page 
transition and corresponds in the field theory to a transition
between the confining and unconfined phases. 
Since a microcanonical framework requires that energy be conserved during any
transition, we shall consider not empty AdS with thermal identifications, 
but rather thermally-identified AdS with a thermal gas in it. The energy 
of the system is then measured with respect to the thermal AdS background.

This paper is structured as follows. We begin, in Section II, by reviewing the
thermal properties of AdS black holes and their interpretation in light
of the AdS/CFT correspondence. Then, in Section III, we consider 
a thermal bath in AdS. Finally, in Section IV, we determine the 
necessary conditions for a crossover between the
black hole and radiation-dominated phases. 

\section{AdS Black Holes}

The five-dimensional Einstein-Hilbert action with a cosmological
constant is given by 
\beq  
I_{\rm BH} = -{1 \over 16 \pi G_5} \int d^5 x \sqrt{-g} \lf R + 12 l^2 \rt \; ,
\label{EH}
\eeq 
where $G_5$ is the five-dimensional Newton constant, $R$ is the
Ricci scalar, the cosmological constant is $\Lambda = - 6 l^2$, and we
have neglected a surface term at infinity.  Anti-de Sitter solutions
derived from this action can be embedded in ten-dimensional IIB
supergravity such that the supergravity background is of the form
$AdS_5 \times S^5$.

The line element of a ``Schwarzschild'' black hole in anti-de Sitter
space \cite{hawkingpage} in five spacetime dimensions can be written
as  
\beq 
 ds^2 = - \lf 1 - {2 M G_5 \over r^2} + r^2 l^2 \rt dt^2 +
\lf 1 - {2 M G_5 \over r^2} + r^2 l^2 \rt^{\! \! -1} dr^2  + r^2 d
\Omega^2_3 \; ,  \label{HPds2}
\eeq 
where $l$ is the inverse radius of AdS space.
This solution has a horizon at $r= r_+$ where 
\beq  r_+^2 = {1 \over 2 l^2} \lf -1  + \sqrt{1 + 8M G_5
\, l^2} \rt \; . \label{r+}  
\eeq
When $r_+ l \ll 1$, the black hole could become unstable to localization on
the $S^5$ by an analog of the Gregory-Laflamme mechanism \cite{ruth}. 
As a rule, one may determine a necessary (though not sufficient) condition for
instability from entropic considerations. 
A straightforward computation then shows 
that localization instability could occur 
for very small black holes with $r_+ l \ll 1$ \cite{gary}. 
Here we shall work with black holes
with $r_+ l > 1$ for which we do not expect such an instability.

To study the black hole's thermodynamics, we Euclideanize the 
metric. The substitution $\tau = i t$
makes the metric positive definite and, by the usual removal of the
conical singularity at $r_+$, yields a periodicity in $\tau$ of 
\beq
\beta_{\rm BH} = {2 \pi r_+ \over 1 + 2r_+^2 l^2} \; ,\label{invtemp}
\eeq
which is identified with the inverse temperature of the black
hole. The entropy is given by   
\beq  
S_{\rm BH} = {A \over 4 G_5} = {\pi^2 r_+^3 \over 2 G_5} \; , \label{S}   
\eeq  
where $A$ is the ``area'' (that is, three-volume) of the horizon.
The mass above the anti-de Sitter background is  
\beq 
U_{\rm BH} 
= {3 \pi \over 4} M = {3 \pi \over 8 G_5} r_+^2 \lf 1 + r_+^2 l^2 \rt  \;
. \label{UBH} 
\eeq 
This is the AdS equivalent of the ADM mass, or
energy at infinity. (Actually if the black hole is to be considered at thermal
equilibrium it should properly be regarded as being surrounded by a thermal
envelope of Hawking particles. Because of the infinite blueshift at the
horizon, the envelope contributes a formally infinite energy. 
Here we shall neglect this infinite energy as unphysical, absorbed perhaps
by a renormalization of the Newton constant.) 
We can now also write down the free energy:
\beq
F_{\rm BH} = {\pi r_+^2 \over 8 G_5} \lf 1 - r_+^2 l^2 \rt \; . \label{Fbh}
\eeq
Eqs. (\ref{S}-\ref{Fbh}) then satisfy the first law of thermodynamics.
To express them in terms of the gauge theory parameters 
$N$, $T_{\rm CFT}$, and $V_{\rm CFT}$, we substitute physical data taken from 
the boundary of the black hole
spacetime. At fixed $r \equiv r_0 \gg r_+$, the boundary line element
tends to  
\beq  
ds^2 \to r_0^2 \left [ - l^2 dt^2 + d \Omega_3^2 \right ] \; ,
\eeq 
giving a volume of
\beq  V_{\rm CFT} = 2 \pi^2
r_0^3 \; .   
\eeq 
The field theory temperature is the physical temperature at the boundary:
\beq
T_{\rm CFT} = {T_{\rm BH} \over \sqrt{-g_{tt}}} \approx 
{T_{\rm BH} \over l r_0} \; .
\eeq
To obtain an expression for $N$, we invoke the
AdS/CFT correspondence. This 
relates $N$ to the radius of $S^5$ and the cosmological
constant: 
\beq  R^2_{S^5} = \sqrt{4 \pi g_s {\a '}^2 N} = {1 \over
l^2} \; .  
\eeq 
Then, since
\beq
(2 \pi)^7 g_s^2 {\a '}^4 = 16 \pi G_{10} =  16 {\pi^4 \over l^5} G_5 \; , 
\eeq 
we have 
\beq  
N^2 = {\pi\over 2 l^3 G_5} \; . \label{N} 
\eeq
With these substitutions, we see that in the limit $r_+ l \gg 1$, 
the black hole entropy can be expressed in terms of conformal 
field theory parameters as
\beq
S_{\rm BH} =  {2 \over 3} \pi^2 N^2 V_{\rm CFT} T_{\rm CFT} ^3 \; .
\eeq
The dimensionful terms in this expression are 
in accord with expectations for a conformal theory.
The matching has been extended \cite{5dadskerr,holorotbh,harvey} 
to rotating black holes and
their field theory dual, Yang-Mills with angular momentum.
The dependence on $N^2$ indicates that the conformal field theory is in its
unconfined phase; the $N^2$ species of 
free gluons make independent contributions to the free energy. We shall
see in the next section that the thermodynamics of the confined phase 
is rather different.

\section{A hot bath in AdS}

Now consider a gas of thermal radiation in anti-de Sitter space. The 
energy eigenstates of $AdS_5$ are \cite{isham}:
\beq
\Psi_{\omega j m n} (r, t, \theta, \phi, \psi) = 
N_{\omega j} \, \exp \lf -i \, \omega \, l \, t \rt \,
\sin ^j \rho \,
C^{j+1}_{\omega -j-1}(\cos\rho) \, Y^{m n }_j (\theta, \phi, \psi) \; ,
\eeq
with the condition $\omega - 1 \geq j \geq |m|,|n|$ where $C^{p}_{q} \, (x)$
are Gegenbauer polynomials, $Y^{m n}_j (\theta, \phi, \psi)$ 
are the spherical harmonics in five-dimensional
spacetime (with total angular momentum number $j$), and 
$\rho \equiv \arctan (rl)$. Here $\omega$ is an integer and  
hence the spectrum is quantized in units of $l$, 
the inverse ``radius'' of AdS. 
Since this is also the quantum of excitations of the five-sphere, 
we should consider thermodynamics over the full ten-dimensional space.
The appropriate line element is therefore
\beq 
ds^2 = - \lf 1 + r^2 l^2 \rt dt^2 + \lf 1 + r^2 l^2 \rt^{-1} dr^2 + r^2 d
\Omega_3^2 + l^2 d \Omega_5^2 \; . \label{AdS}
\eeq 
To obtain a thermal field theory, we again Euclideanize the metric. The
periodicity of $\tau = it$ is then the inverse 
(asymptotic) temperature, $T^{-1}_{\rm AdS}$, of the theory; the absence of
a horizon means that $T_{\rm AdS}$ is an arbitrary parameter. However, the
relevant temperature for thermodynamics in the bulk is not $T_{\rm AdS}$, but
the local, redshifted, temperature:
\beq 
T_{\rm local} = {T_{\rm AdS} \over \sqrt{-g_{tt}}} 
= {T_{\rm AdS} \over \sqrt{ 1+ r^2 l^2}} \; . \label{redshift}
\eeq
To calculate thermodynamic quantities we foliate spacetime into (timelike)
slices of constant local temperature. Extensive thermodynamic quantities
are then computed by adding the contribution of each such hypersurface.
 
The local five-dimensional energy density of the thermal gas of 
radiation can be written as
\beq 
\rho_{\rm local} = \sigma {\pi^3 \over l^5} T^{10}_{\rm local} \; , \label{rho}
\eeq 
where we have neglected infrared effects due to curvature or nonconformality.
Here $\sigma$ is the ten-dimensional supersymmetric 
generalization of the Stefan-Boltzmann constant, which is approximated
by its flat space value:
\beq 
\sigma = {62 \over 105} \pi^5 \; ,
\eeq
where we have included a factor of 128, the number of 
massless bosonic physical degrees of freedom of IIB supergravity.

The total ``ADM'' energy-at-infinity of a gas contained in a ball
of radius $r_0$ is then
\beq
U_{\rm gas}^{\infty} 
= \sigma {\pi^3 \over l^4} \int T^{10}_{\rm local} 
\sqrt{-g_{tt}} \sqrt{g_{rr}} \,
r^3 dr \, d \Omega_3 = {2 \pi^5 \over l^5} \sigma T_{\rm AdS}^{10} 
\int_0^{r_0} {r^3 dr \over
{(1+r^2 l^2)^5}} \equiv \sigma V_{\rm eff} (r_0) T_{\rm AdS}^{10} \; . 
\label{Ugas} 
\eeq
Here the additional blueshift factor of $\sqrt{-g_{tt}}$
converts the local (fiducial) energy into an ADM-type 
energy, comparable to Eq. (\ref{UBH}). We have also defined an 
effective volume,
\beq 
V_{\rm eff}(r_0)= {2 \pi^5 \over l^9} \lf {2 \over 3} -{ {2+ 3
(r_0 l)^2 } \over {3 \lf 1 + (r_0 l)^2 \rt ^{3/2} }} \rt  \; , 
\eeq
which, as $r_0 \rightarrow \infty$, approaches
\beq 
{4 \pi^5 \over {3l^9}} \; .
\eeq
Thermodynamically, anti-de Sitter space behaves as if it had a
finite volume.

Similarly, the other thermodynamic quantities of the thermal bath are
\beq 
F = - {\sigma \over 9} V_{\rm eff} T_{\rm AdS}^{10} \comma 
S = {10 \over 9} \sigma V_{\rm eff} T_{\rm AdS}^9 \; ,
\eeq
consistent with the first law of thermodynamics.
The absence of a $G_5$  in the free energy
indicates, from the CFT point of view, 
that the free energy is of order $N^0$. This is the confined
phase of the theory -- the free energy is of order $N^0$ because the $N^2$
species of gluons have condensed into hadronic color singlets. (Curiously,
attempts to formulate confinement in terms of anti-de Sitter space date
back at least to the 1970's \cite{salam}.)

The factor of nine spatial dimensions in the volume is somewhat puzzling for
a three-dimensional gauge theory. It reflects the fact that the 
QCD (SYM) string is really a type IIB string which naturally lives in nine 
spatial dimensions. It has been suggested that
the extra dimensions in which the open string worldsheet 
bounded by a Wilson loop can extend are akin to 
Liouville dimensions \cite{cave,adscft,juanwilson,nadavgross}. 

\section{The Crossover}

A field theory living on a
manifold, $S^3 \times S^1$ with three-sphere radius $r_0$, is dual to
those five-dimensional Einstein manifolds that have the same geometry
at $r_0$. In the microcanonical approach that we shall follow, the
contributions to the partition function come from both the black
hole and the gas in thermally-identified AdS where the energy of the 
gas and black hole are taken to be the same:
\beq
Z(U) = e^{-I_{\rm BH}(U)} + e^{-I_{\rm gas}(U)} \; . \label{part}
\eeq
Which of these two thermodynamic phases the system is found in is
determined, in the saddle point approximation, by the relative values of
the respective Euclidean classical actions. The action of the black hole is
simply the Einstein-Hilbert action, Eq. 
(\ref{EH}). This is proportional to the volume of the spacetime and so
needs to be regulated. A finite action is obtained by subtracting the
(also infinite) action for thermally-identified 
anti-de Sitter space in which the
hypersurface at a constant large radius has the same intrinsic 
geometry as a hypersurface at the same radius in the black hole background
\cite{hawkingpage}. The regularized black hole action is
\beq
I_{\rm BH} = {\pi^2 r_+^3 \over 4 G_5} 
{1 - r_+^2 l^2 \over 1 + 2 r_+^2 l^2} \; . 
\label{IBH}
\eeq
Subtracting the anti-de Sitter background is equivalent to choosing the
ground state of the theory. Then the comparable value of the action for 
the gas in thermally-identified AdS should be just the action of the 
gas itself, namely $F/T$. Thus
\beq
I_{\rm gas} = -{1 \over 9} \sigma V_{\rm eff} T_{\rm AdS}^{9} \; . \label{Igas}
\eeq

The qualitative thermodynamic behavior of the system is determined by
the action which dominates the partition function Eq. (\ref{part}). At the
crossover between the two phases, the action for the
gas and the black hole are the same. Moreover, energy must be conserved.
Hence one may determine the conditions for a smooth crossover from the
following equations:
\beq
U^{\rm local}_{\rm gas} = U^{\rm local}_{\rm BH} \comma
I_{\rm gas} = I_{\rm BH} \; .
\eeq
Note that, since the two phases cannot be in physical contact, 
the physical temperature does not have to be the same for the two
phases. (The physical temperatures are the same at $r_+l = \sqrt{7}$.)

Solving these equations yields $N^2$ as a function of the dimensionless
quantity $x \equiv r_+ l$ at the crossover:
\beq
N^2 = {31 \over 2^5 \cdot 3^{13} \cdot 5 \cdot 7} \, 
{(1+x^2)^9 \cdot (1+2x^2)^{10} \over (x^2 - 1)^{10} \cdot x^{12}} \; . 
\label{cross}
\eeq

How accurate is this equation? In using Eq. (\ref{redshift}), 
we have omitted the back-reaction of the gas on the
metric. One may estimate this. Consider a
spherically symmetric stationary spacetime with a cosmological
constant and massive matter fields. The equation of hydrostatic
equilibrium (equally, the $t-t$ Einstein equation) reads
\beq 
{d \over dr} m(r) = 8 \pi^2 r^3 \lf G_5 \rho + \Lambda \rt \; ,
\eeq
where $- g_{tt} \equiv 1 - m(r) / r^2$. Back-reaction can reliably be
neglected when the matter term in the parenthesis is (much) smaller than the
cosmological term. For an energy density given by Eq. (\ref{rho}) and
a total energy matched to that of the black hole phase, Eq. (\ref{UBH}),
the condition $G_5 \rho < |\Lambda|$ amounts to
\beq
{9 \pi \over 32} \, x^2 (1 + x^2) \, l^2 < 6 \, l^2 \; ,
\eeq
and we see that the matter term becomes dominant at large $x$, and is not
entirely negligible even near $x = 1$. At high temperature, therefore,
Eq. (\ref{cross}) becomes unreliable; this limit has been studied elsewhere
\cite{joserabin}. To accommodate the effect of matter, one might try
to seek a solution to the linearized Einstein equations, 
perturbed around the $AdS_5$ background.
The exact form of Eq. (\ref{cross}) would also be modified by 
including the correct Stefan-Boltzmann constant for anti-de Sitter space.
And finally, when $N$ is small,
the supergravity approximation itself breaks down.

\begin{figure} \begin{center} \includegraphics[angle=-90,
width=120mm]{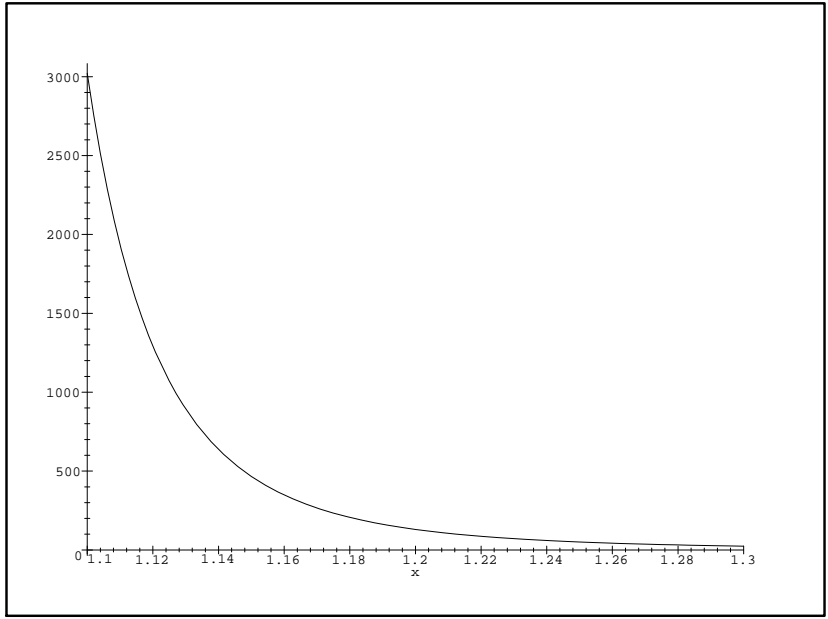} \caption{$N$ vs. $x \equiv r_+ l$ for $x$
near unity.} \end{center} \label{graph}
\end{figure}
Despite these caveats, Eq. (\ref{cross}) seems to capture the correct
qualitative behavior. In Fig. 1, we plot $N$ near the crossover for 
$x \sim 1$. The region below the crossover curve is dominated by the confined 
or AdS gas phase, whereas the region above is dominated by 
the unconfined or black hole phase. 
Note that $x$ grows roughly like the dimensionless 
product $T_{\rm phys} \, r_0$. As the temperature/volume increases, the
graph confirms our expectation that the theory becomes conformal.
As $N$ goes to infinity, we recover the result \cite{thermalphase}
that the transition occurs at $r_+ l = 1$. This is in fact
a very good approximation for finite but large N.

\textsc{Acknowledgments}

We would like to thank Jos\'e Barb\'on, Gerard 't Hooft, Shiraz Minwalla,
and Erik Verlinde for helpful
discussions. D. B. is supported by
European Commission TMR programme ERBFMRX-CT96-0045, and 
would like to acknowledge the hospitality of
the Spinoza Institute, CERN and the support of ITF, Utrecht. M. P. is
by the Netherlands Organization for Scientific Research (NWO).

\end{document}